\documentstyle[prl,aps,multicol,epsf]{revtex}
\input epsf
\date{Received July 28, 2001}

\begin{document}
\draft
\title{Adsorption in non interconnected pores open at one or at both
ends:\\
a reconsideration of the origin of the hysteresis phenomenon.}
\author{B. Coasne, A. Grosman, C. Ortega and M. Simon}
\address{Groupe de Physique des Solides, Universit\'es Paris 7 et 6,
   UMR-CNRS 75-88, 2 place Jussieu, 75251 Paris Cedex, France.}

\maketitle

\begin{abstract}
We report on an experimental study of adsorption
isotherm of nitrogen onto porous silicon with non interconnected
pores open at one or at
both ends in order to check for the first time
the old (1938) but always current idea
based on Cohan's description which suggests that the adsorption
of gas should occur reversibly in the first case and irreversibly
in the second one.  Hysteresis loops, the shape of which is usually
associated to interconnections in porous media, are observed whether
the pores are open at one or at both ends in contradiction with Cohan's model.
\\
\end{abstract}

\begin{multicols}{2}

Adsorption isotherms in mesoporous materials exhibit two
main typical features: (i) a sharp increase in the amount
of gas adsorbed at pressure below the saturated vapor
pressure of the bulk gas which is attributed to capillary
condensation in pores of confining sizes; (ii) the desorption
process fails to retrace the path of the adsorption one, so that
a hysteresis loop appears revealing the irreversibility of the
phenomenon.

The first explanation of the capillary condensation was proposed
by Zsigmondy \cite{Zsigmondy} who, for this purpose, refered to the macroscopic
Kelvin equation. In order to explain the hysteresis phenomenon, Cohan
\cite{Cohan} went further and proposed different scenari for filling and
emptying of a pore which he also described with the macroscopic
concept of meniscus. In a cylindrical pore open at both ends,
the shape of gas/adsorbate interface is different during the
adsorption and desorption processes leading to a hysteresis
phenomenon. On the contrary, in a cylindrical pore closed at one end, the
adsorption is reversible since the same meniscus is present during the
filling and the emptying of the pore.

The irreversibility of the adsorption process in a cylindrical pore
open at both
ends, expected by Cohan on the basis of thermodynamic
considerations, has been predicted by new numerical approaches.
Thus, Grand Canonical Monte-Carlo (GCMC) simulations
\cite{Papadopoulou}, analytical
approaches \cite{Celestini} together with Density Functional Theory (DFT)
\cite{Evans} have also proposed that the hysteresis phenomenon is an intrinsic
property of the phase transition in a single idealized pore and arises
from the existence of metastable gas-like and/or liquid-like states.
It is noteworthy
that no calculations have been yet performed for cylindrical pores closed
at one end.

Macroscopic descriptions based on Cohan's model are currently used to
describe the adsorption in non interconnected mesoporous media and in
particular to
determine the Pore Size Distribution (PSD) via the
Barrett-Joyner-Halenda (BJH) method \cite{BJH}.

However, the experimental validation of Cohan's model has never
been checked since
the synthesis of ordered mesoporous media with non interconnected
pores open at one or at both ends was unsuccesful for a long time.
Using porous silicon, we report in this paper, for the first time, on an
experimental investigation of Cohan's predictions. For this check,
this mesoporous material exhibits two essential properties: (i) the
tubular pores are non interconnected and do not exhibit narrow
sections and (ii) it exists under two forms
with pores closed at one end when the porous layer is supported
by the substrate or open at both ends when the layer is removed from
the substrate.

Porous silicon is prepared by electrochemical etching of Si
single crystal in a HF solution. The morphology of the porous Si
layers depends on the type and concentration of the dopant
atoms \cite{GrosmanOrtega1}. The adsorption studies presented here have
been performed with porous layers prepared from highly boron doped
[100] Si substrate. The porosity of the layers is controlled
by the current density and HF
concentration, and the thickness by the duration of the anodic
dissolution. The pore volume is determined with high precision by
weighting the sample before and after the electrochemical dissolution.
To calibrate the porosity and the thickness of the layer we need a
third weighing after the dissolution of the porous part in a
NaOH solution. It is noteworthy that the porous
layers are perfectly reproduced. The porous layers under study
had a porosity of 51\% and the length of the pores is
21 ${\rm \mu}$m.\\ The morphology of these layers (see
Fig.~\ref{TEM}(a) and \ref{TEM}(b)) can be described as honeycomb-like
structure with pores perpendicular to the Si substrate separated
from each other by Si single crystal walls (inset Fig.~\ref{TEM}(b))
of constant thickness ($\sim$ 5 nm).

The pores, with sections of polygonal shape, are of the same length and
closed at one end by the Si substrate. At the end of the porous
layer formation which takes about 20 minutes, an increase of the 
current density
leads to the electropolishing regime during which the Si walls at
the bottom of the pores are dissolved so that the porous layer comes off the
Si substrate. It is noteworthy that such a process takes only a few
seconds and thus that the PSD of the porous layer is not changed.
Playing with this possibility, porous Si layer or
porous Si membrane can be obtained with pores open at one or at both
ends, respectively.\\
\begin{minipage}{8.3cm}
\begin{figure}
\begin{center}
\epsfxsize=6cm
\epsfbox{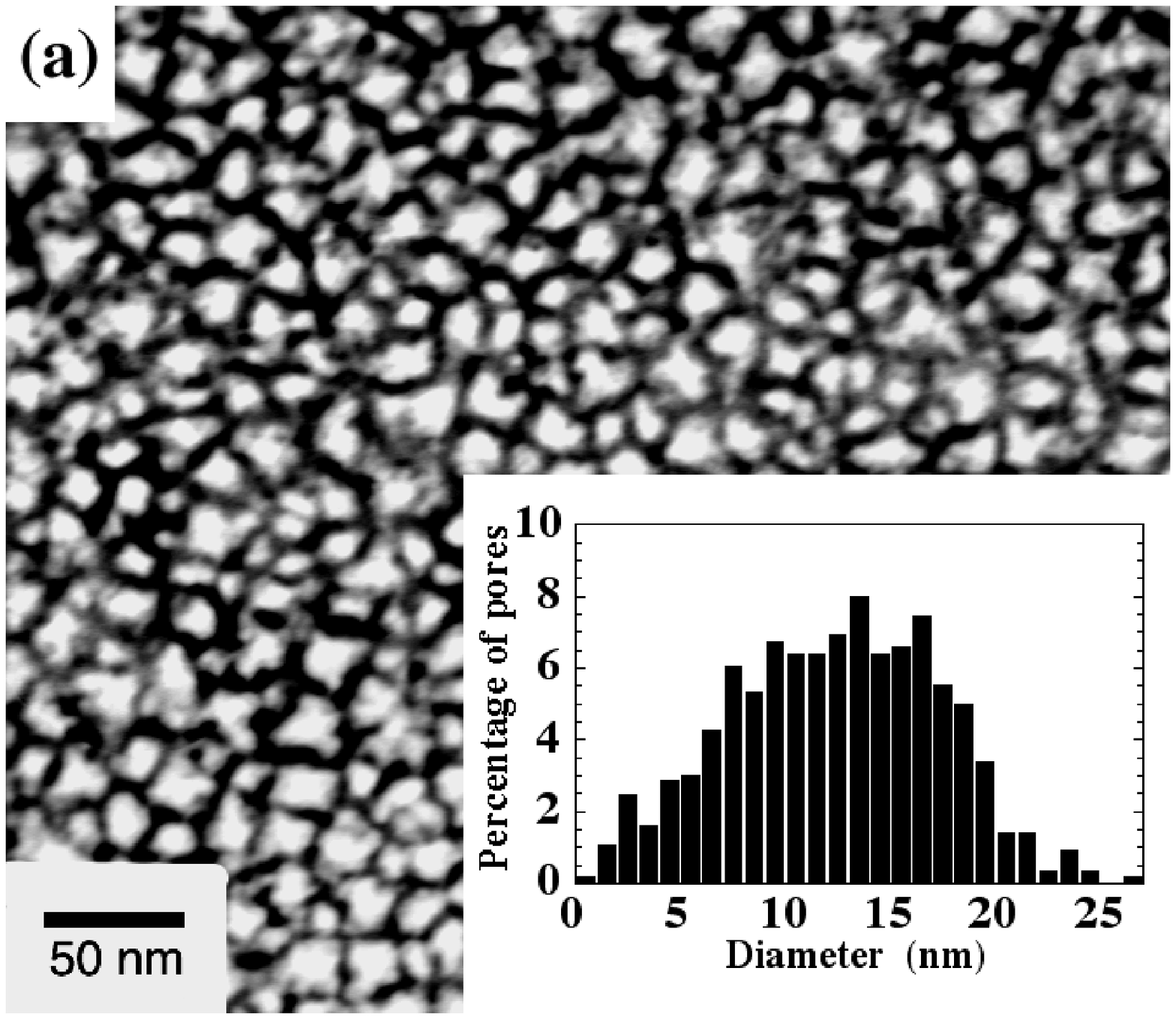}
\epsfxsize=6cm
\epsfbox{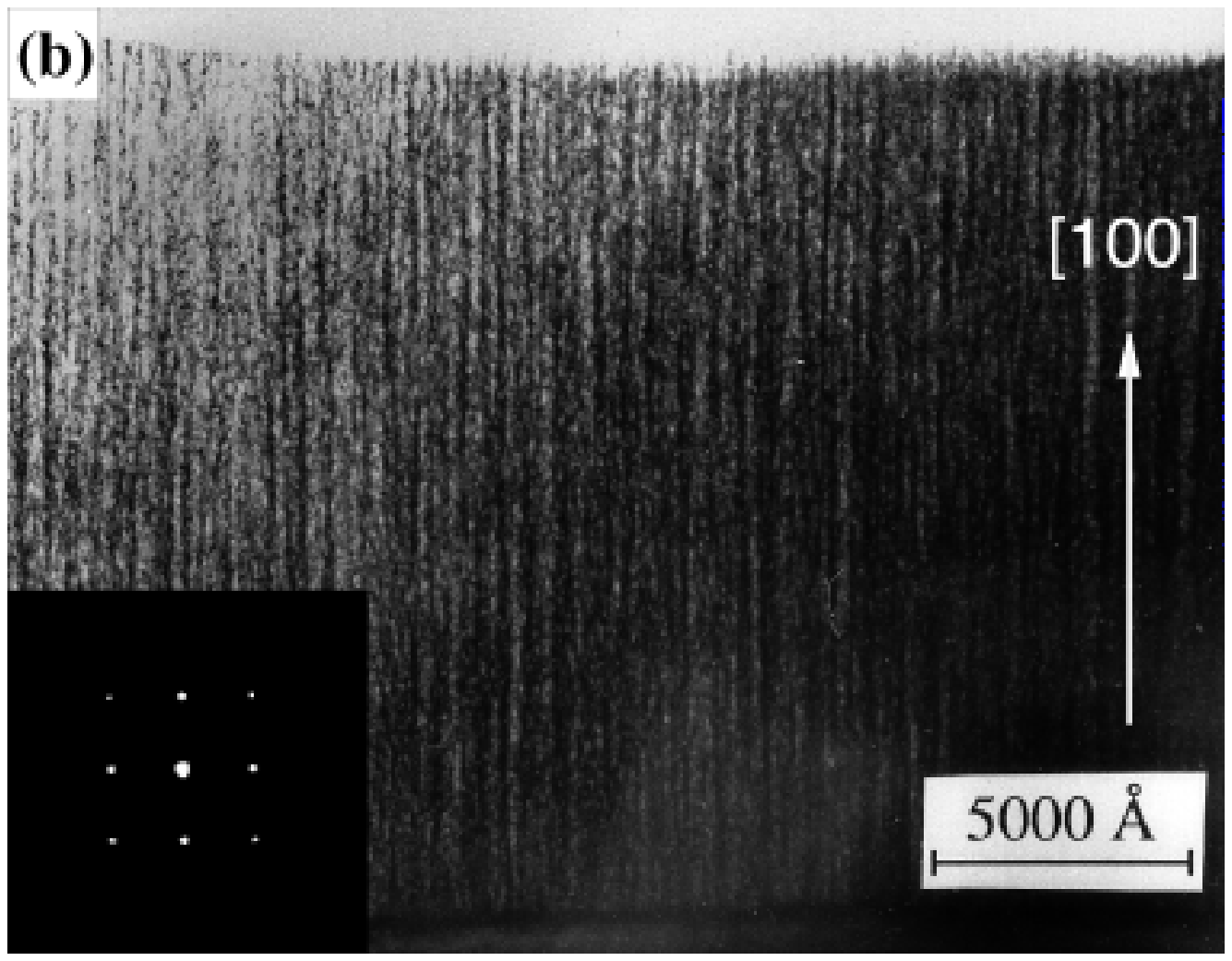}
\vskip2mm
\caption{Bright field TEM images in plane view (a) and cross-section
(b) of a porous Si
layer with 51\% porosity prepared from $p^{+}$ type ($\sim
3.10^{-3} \Omega$.cm)
[100] Si substrate. Observation axis [100]. Pores (white) are separated by
Si walls (black). The pore density is $3.2.10^{11}/{\rm cm^{2}}$. Inset
1a: PSD. Inset 1b: transmission electronic diffraction pattern.}
\label{TEM}
\end{center}
\end{figure}
\end{minipage}
\vskip2mm
We have shown that there is no lateral interconnection between the
pores in these layers. The method involves deposition
by evaporation at ambient temperature, of an aluminum layer, 500 nm thick,
on a part (${\rm A_{c}}$ and ${\rm B_{c}}$) of the top of two porous
samples A and B,
following by a thermal annealing under neutral atmosphere of argon at
$450^{o}$C for
half a hour to improve the contact between the Al deposit and the top of
the layers. We have shown by Rutherford BackScattering analysis (RBS)
with 2~MeV $\alpha$ particles that, the Al deposited on
the samples
is well localized on the top of the pores and constitutes a cap. One
of the two samples (A) was thermally oxidized in ${\rm
O}_{2}$ enriched in $^{18}{\rm O}$ (99\%), at $300^{o}$C,  12 mbar,
for one hour.
This treatment leads to the
formation of a thin silicon oxide ($\sim$1~nm) on the silicon walls.
The $^{18}{\rm O}$ contents in the porous layers under the Al cap (${\rm
A_{c}}$ and ${\rm B_{c}}$)
and beside it (${\rm A_{o}}$) were determined by the nuclear reaction
$^{18}{\rm O}$(p,$\alpha$)$^{15}{\rm N}$
at a proton energy near the narrow resonance at ${\rm E_{p}}$=~629~keV
($\Gamma$~=~2~keV).
For the measurements under the Al caps the energy of the protons beam
energy was increased to take into account the energy loss of the
incident protons in the Al layer. The nuclear reaction spectra corresponding
to ${\rm A_{o}}$ and ${\rm A_{c}}$ are shown in Fig.~\ref{NRA}.
The $\alpha$ peak from ${\rm A_{o}}$ indicates of the presence in the 
porous layer of $^{18}{\rm O}$ atoms coming
from the thermal oxidation. Due to the presence of Al cap, the $\alpha$
peak from ${\rm A_{c}}$ is expected at an energy lower by 100~keV than
that from the $\alpha$ peak from ${\rm A_{o}}$. The counts from ${\rm
A_{c}}$ and ${\rm B_{c}}$ which are one order of magnitude
lower than
that from ${\rm A_{o}}$ are equal,
and hence corresponds only to a background coming from nuclear
reactions with silicon, boron atoms, etc.
This experiment reveals that this
medium is composed of non interconnected single pores,
the top of which is in direct contact with the gas reservoir
during adsorption. To our knowledge, this is the only porous material
for which the absence of interconnection has been experimentally
proven.\\
\begin{minipage}{8.3cm}
\begin{figure}
\epsfxsize=8cm
\epsffile{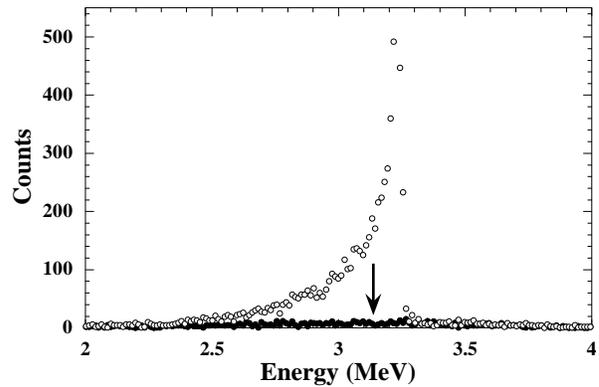}
\vskip1mm
\caption{$\alpha$ spectra from $^{18}{\rm O}$(p,$\alpha$)$^{15}{\rm
N}$: ${\rm E_{p}}$=~629~keV. White circles: porous Si layer after thermal
oxidation in $^{18}{\rm O}_{2}$ (${\rm A_{o}}$). Black circles:
part of the layer protected by Al cap (${\rm A_{c}}$). The arrow
indicates the energy of the expected $\alpha$ peak from ${\rm A_{c}}$.}
\label{NRA}
\end{figure}
\end{minipage}
\vskip2mm
Another important property of this medium is the constancy
of the Si walls thickness
revealed by the TEM plane view performed on a porous layer thinned
down to 1000 \AA, i.e. about 10 times the size of the pores.
This is a strong argument against the presence of narrow sections inside the
pores, the presence of which would lead, through
shadow effects, to an inhomogeneous apparent thickness of the Si
walls.

The PSD of the porous Si layers have been estimated from a
numerical treatment of the TEM images. The negative TEM photographs
are digitized by means of a CCD monochrome camera. Then, the
value of the threshold to make the image binary is chosen to
reproduce the porosity of the sample. The analysis of this binary
image allows one to obtain the cross section area and perimeter of each pore.
The PSD, which corresponds to cylindrical pores having the same
surface area as the polygonal pores, is shown in inset of
Fig.~\ref{TEM}(a). The mean diameter of the pores is 13nm with a
standard deviation of $\pm$6nm.

Adsorption measurements were made by a volumetric technique using
a commercial apparatus under secondary
vacuum of around $10^{-6}$ torr. Nitrogen adsorption isotherms were
measured up to the saturating vapor pressure of bulk
nitrogen (P$_{0}$) and at 77 K. Porous Si
layers were introduced in a cell just after
their preparation. The samples were outgassed at ambient
temperature for a few hours. This outgassing does not change the nearly
complete passivation of the inner surface by a monolayer
of ${\rm SiH_{1,2,3}}$ groups which takes place during the
electrochemical etching
and which is stable for a few days after the formation
\cite{GrosmanOrtega1}.\\
\begin{minipage}{8.3cm}
\begin{figure}
\epsfxsize=8cm
\epsffile{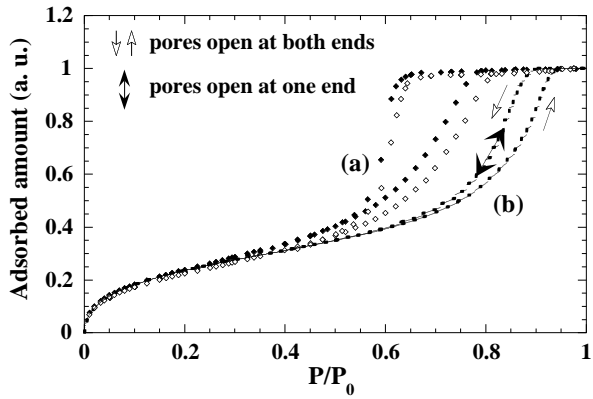}
\vskip1mm
\caption{(a) N$_{2}$ adsorption isotherms at 77K for porous Si with
pores open at one (black diamonds) or at both (white diamonds) ends.
(b) Isotherms calculated by introducing in Eqs.~(\ref{Cohanadsorption})
and (\ref{Cohandesorption}) the experimental t(${\rm P}$) values (see
text) and the PSD
shown in Fig.~\ref{TEM}.}
\label{isothermes}
\end{figure}
\end{minipage}
\vskip2mm

Figure~\ref{isothermes}(a) shows N$_{2}$ adsorption
isotherms at 77 K corresponding to porous Si layers with pores open at
one or at both ends, the PSD of which is represented
in inset of Fig.~\ref{TEM}(a). These two curves exhibit wide and asymmetrical
hysteresis loops with shapes corresponding to type H2 with reference
to the 1985 IUPAC classification (or type E in that of de Boer
\cite{DeBoer}).

The adsorbed amount of gas sharply increases between 0.55 and 0.8 P$_{0}$
and then reaches
a plateau region where all the pores are filled with the dense phase. The
steep desorption process occurs over a pressure range about three
times narrower than that of the adsorption.

On the assumption that the density of the dense phase
equals that of the bulk liquid nitrogen, we found that
the pore volume extracted from the value of the adsorbed
amount on the plateau region is in a very good agreement
with that measured by gravimetry.

Cohan's model predicts that, for a cylindrical pore open at
both ends, the condensation takes place at a pressure
${\rm P_{c}}$ given by an equation similar to the classical Kelvin equation
where the curvature of the meniscus is replaced by a cylindrical
curvature :
\begin {equation}
\ln\left ( \frac{{\rm P_{c}}}{{\rm P}_{0}}\right ) = - \frac{\sigma {\rm
V_{L}}}{\rm RT(r-t(P_{c}))}
\label {Cohanadsorption}
\end {equation}
where V$_{\rm L}$ is the molar liquid volume, $\sigma$ is the
liquid-gas surface tension at temperature T, r is the radius of
the pore and t(${\rm P}$) the thickness of the adsorbed layer at P.
In the case of a cylindrical pore closed at one end, surface
thermodynamics
considerations predict a preferential adsorption on the bottom of the
pore; a hemispherical meniscus is then present provided that there is
perfect wetting,
so that the condensation occurs
at a pressure ${\rm P_{h}}$ given
by the following modified Kelvin equation:
\begin {equation}
\ln\left (\frac{{\rm P_{h}}}{{\rm P}_{0}}\right ) = - \frac{2 \sigma {\rm
V}_{L}}{\rm RT(r-t(P_{h}))}
\label {Cohandesorption}
\end {equation}
The evaporation of the wetting fluid proceeds through the same
hemispherical meniscus in both cases i.e. at the same pressure
${\rm P_{h}}$.

Thus, according to Cohan's model : (i) the adsorption
phenomenon should be reversible in pore open at one end while it should be
irreversible in pore open at both ends; (ii) the adsorption branch for a
pore open at both ends should be located at a pressure (${\rm P_{c}}$)
higher than that for a pore closed at one end (${\rm P_{h}}$);
(iii) for a given size distribution of
cylindrical pores open at both ends, the adsorption
branch should be steeper than that of the
desorption branch (provided that the diameter exceeds 3-4 nm)
and the shape of the hysteresis loop becomes rather symmetrical
as the pore diameter
increases, as shown by numerical calculations based on
Eqs.~(\ref{Cohanadsorption}) and (\ref{Cohandesorption}) in which we
have introduced a currently used t(${\rm P}$) proposed by Halsey
\cite{Halsey,Gregg} or the t(${\rm P}$) reproducing our experimental values
prior to the capillary condensation.

To extract the latter t(${\rm P}$) values, we have prepared a porous Si layer
with a PSD having a mean value as high as possible (about 50 nm)
in order to shift the capillary
condensation pressures to the highest possible values. The obtained
isotherm exhibits a hysteresis loop located above $0.8{\rm P}_{0}$.
We have thus determined the numerical
values of t(${\rm P}$) up to $0.8{\rm P}_{0}$ by distributing homogeneously
at each pressure the adsorbed amount among the pores, the PSD of which
was extracted as explained above.
In order to draw the t(${\rm P}$) above $0.8{\rm P}_{0}$, we have 
chosen to extrapolate
the polynomial law which fits the experimental t(${\rm P}$).

The isotherm calculated by introducing in
Eqs.~(\ref{Cohanadsorption}) and (\ref{Cohandesorption}) the t(${\rm P}$)
and the PSD shown in inset Fig.~\ref{TEM}(a) is represented
Fig.~\ref{isothermes}(b).

We will successively discuss the predictions (i) and (iii) which are in large
discrepancy with the experimental results (Fig.~\ref{isothermes}(a));
the prediction (ii) which is in agreement with experiment
and finally, the large shift between the absolute position of the
experimental isotherms and the calculated ones.

(i) The adsorption-desorption cycle observed in the case of pores
closed at one end is irreversible. The interconnectivity related
pore blocking effect
described by Mason \cite{Mason} to explain the origin of the hysteresis
phenomenon could not occur because the porous Si layers are composed
of non interconnected pores. Moreover, we reject the `ink bottle' effect,
previously introduced by Everett \cite{Everett}, as a possible explanation
of the hysteresis loop we observe on the adsorption isotherms because
of the absence of narrow sections inside the pores.

(iii) The hysteresis loops observed in the porous layer and porous
membrane exhibit the same shape of type H2. We have observed, by
carrying out adsorption isotherms on porous Si
layers with a PSD having mean values varying from 13 nm up to 50 nm,
that the position on the pressure axis of the corresponding
adsorption and desorption branches are shifted towards ${\rm P}_{0}$
whereas the shape of the hysteresis loops remains of type H2.

Such an asymmetrical shape
is not expected by the single pore models. Indeed,
numerical calculations based on Cohan's model or
on DFT \cite{BallEvans} predict a symmetrical shape of the hysteresis
loop for such large pores in
contradiction with what we observe. Furthermore, although porous
Si exhibits non-interconnected pores, the hysteresis shape
(type H2) we observe is the same as those systematically obtained
for highly interconnected mesoporous solids such as controlled
pore glass or Vycor \cite{Brown}. From a theoretical point of view, both
macroscopic \cite{Mason} and microscopic \cite{BallEvans} approaches
described this
shape as a signature of the disorder introduced by the
interconnectivity between pores. We must conclude that a
hysteresis loop of type H2 is not characteristic of
interconnected porous materials and we thus reject a previous
conclusion made by Ball and Evans according to which a de Boer's
shape of hysteresis is a consequence of interconnected network effect.

We believe that the desorption process we observe results from another
effect. At the end of the adsorption process, when all the pores are
filled by the dense phase, the external surface of the porous layer
is covered by a film which connects the pores. A possible interpretation
of the observed delay to the emptying of pores could be the absence of
formation of a concave meniscus on top of each pore. Thus, question
arises whether metastable states exist which would correspond to a flat
interface between the film and gas. In such an idea, during the desorption
process, the film thickness could decrease homogeneously until the system
jumps to the stable state corresponding to a concave meniscus on top of
each pore. The steepness of the desorption branch could be explained if
the jump is sufficiently delayed until a pressure at which the emptying
of the major part of pores occurs. This idea needs of course to be checked
by numerical calculations.

Note that the analysis of
the desorption branch using BJH method yields a PSD with a
width 0.5~nm about which is very far from that of the actual PSD
($\pm$ 6nm). It is conventionally admitted
\cite{SaamCole} that
the desorption process in porous materials occurs at coexistence
of the high and low dense phases. On such an assumption, a steep desorption
branch is interpretated as the signature of the presence of pores of
unique size\cite{Findenegg}. Our results obviously show that it is 
not necessarily the case.

(ii) The sign of the relative positions
of the two adsorption branches (${\rm P_{h}} < {\rm P_{c}}$)
is in agreement with Cohan's model. These
experiments reveal two different scenari for the
filling of pores open at one or at both ends. Preferential adsorption
in bottom of closed pores is confirmed by analytical calculations
performed in an oblique corner formed by the
intersection of two planar surfaces \cite{Cheng}. We are 
investigating pore closure
effect on adsorption in cylindrical geometry by means of GCMC simulations
and DFT calculations.

As shown in Fig.~\ref{isothermes}(b), the curve which corresponds to
the modified Kelvin equation (2) is largely outside of the experimental
hysteresis loops. For pore sizes as large
as in our samples, this equation should account for the equilibrium
phase transition. Indeed, microscopic calculations based on DFT performed by
Ball and Evans \cite{BallEvans} have shown that, in the case of a 
distribution of
cylindrical pores open at both ends centered on 7.2nm, which is twice smaller
than the mean size of our pores, the coexistence curve is located at
a pressure which is close to that predicted by equation (2).
To our knowledge, it is the first time that,
for so large pore sizes, such a large discrepancy is
found \cite{gubbins}. This may be due to the use of a cylindrical 
geometry in the calculations
which do not account for the actual polygonal shape of pores.
GCMC simulations and calculations based on the DFT are in progress.

In summary, we have studied adsorption phenomenon in
porous Si medium, a well ordered material
with tubular pores without narrow sections and for which we have
experimentally proven the absence of interconnection.
Hysteresis loops of type H2 are observed, and that, whether the pores
are open at one or at both ends. These results are in contradiction
with Cohan's model and show that the only observation of hysteresis
loop of type H2 is not the signature of the presence of
interconnections in porous material as it is generally admitted.
The steep desorption process
cannot be described by the macroscopic Kelvin equation in which we
introduce the experimental values of the thickness of adsorbed amount prior to
the capillary condensation. We note that the application
of the widely used BJH method to extract the PSD from this branch is not
systematically justified.

\end{multicols}


\begin{references}

\bibitem{Zsigmondy}  R.\ Zsigmondy, {\em Z. Anorg. Allgem. Chem.} {\bf
71}, 356 (1911).
\bibitem{Cohan}  L.\ H.\ Cohan, {\em J. Am. Chem. Soc.} {\bf 60},
433 (1938).
\bibitem{Papadopoulou} A.\ Papadopoulou, F.\ Van Swol, U.\ Marini
Bettolo Marconi and P.\ Tarazona, {\em J. Chem. Phys.} {\bf 97}, 6942 (1992).
\bibitem{Celestini} F.\ Celestini, {\em Phys. Lett. A}, {\bf 228} 84 (1997).
\bibitem{Evans} R.\ Evans, U.\ Marini Bettolo Marconi
and P.\ Tarazona, {\em J. Chem. Phys.} {\bf 84}, 2376 (1986).
\bibitem{BJH} E.\ P.\ Barrett, L.\ G.\ Joyner and P.\ H.\ Halenda,
{\em J. Am. Chem. Soc.} {\bf 73}, 373 (1951).
\bibitem {GrosmanOrtega1} A.\ Grosman and C.\ Ortega, in {\em Structural
and optical properties of porous silicon nanostructures}, edited by
G. Amato, C. Delerue and H.J. Von Bardeleben in Gordon and Breach Sc.
Chap.11, pp. 317-332. Chap.13, pp. 375-408 (1997).
\bibitem{DeBoer} J.\ H.\ de Boer, in {\em The structure and
properties of porous materials}, Eds. D. H. Everett and F. S. Stone,
Butterworths, London, p.68 (1958).
\bibitem{Mason} G.\ Mason, {\em J. Colloid Interface Sci.} {\bf 88},
36 (1982).
\bibitem{Everett} D.\ H.\ Everett, in {\em The structure and Properties
of Porous Materials}, edited by D. H. Everett, F. S. Stone,
Proc. Tenth Symp. of Colston Research Society, London, p 95, J.A. Barker
Ibid, p.125 (1958).
\bibitem {Halsey} G.\ Halsey, {\em J. Chem. Phys.} {\bf 16} (10), 931
(1948).
\bibitem{Gregg} S.\ J.\ Gregg and K.\ S.\ W.\ Sing, in
{\em Adsorption, Surface area and Porosity}, Academic Press, London (1982).
\bibitem {BallEvans} R.\ P.\ Ball and R.\ Evans, {\em Langmuir}
   {\bf 5}, 714 (1989).
\bibitem {PCCP} B.\ Coasne {\em et al.}, {\em Phys. Chem. Chem. Phys.} {\bf
3}, 1196 (2001).
\bibitem{Brown} A.\ J.\ Brown, {\em Thesis}, Bristol (1963).
\bibitem{SaamCole} W.\ F.\ Saam and M.\ W.\ Cole,
{\em Phys. Rev. B} {\bf 11}, 1086 (1975).
\bibitem{Findenegg} S.\ Gross and G.\ H.\ Findenegg,
{\em Ber. Bunsenges. Phys. Chem.} {\bf 101}, 1726 (1997).
\bibitem{Cheng} E.\ Cheng and M.\ W.\ Cole,
{\em Phys. Rev. B} {\bf 41}, 9650 (1990).
\bibitem{gubbins}  L. D. \ Gelb, K. E. \ Gubbins, R. \ Radhakrishnan
and M. \ Sliwinska-Bartkowiak, {\em Rep. Prog. Phys.} {\bf
62}, 1573 (1999).
\end{references}
\end{document}